\documentstyle[osa,manuscript]{revtex}  

\begin{document}                

\title{Superconductor-Insulator Transition in Random Two-Dimensional System}

\author{Masahiko Kasuga and Susumu Kurihara }

\address{Department of Physics, Waseda University, 3-4-1 Okubo, Shinjuku-ku, Tokyo 169-8555, Japan} %

\maketitle
\begin{abstract}

Effect of disorder in metallic thin film is examined as a possible mechanism of the Superconductor-Insulator (S-I) transition. The critical value of disorder corresponding to the transition point is found analytically by using Matsubara-Matsuda model and Green's function method. 

\end{abstract}

\section{ INTRODUCTION}
\quad At zero temperature two-dimensional systems of interacting electrons are speculated to show a quantum phase transition between superconducting and insulating (S-I) phases. S-I transition can be driven by tuning some parameters such as the disorder~\cite{rf:Scalettar,rf:Trivedi,rf:Fisher}, carrier concentration~\cite{rf:Phillips,rf:Schon}, magnetic field~\cite{rf:Markovic,rf:Goldman}, dissipation~\cite{rf:Penttila,rf:Yamaguchi}, and so on. It has been observed in granular and amorphous films such as Bi, Pb, Sn~\cite{rf:Scalettar,rf:Trivedi}, Josephson junction arrays~\cite{rf:Penttila,rf:Yamaguchi,rf:Fazio}, and $^4$He absorbed on porous media~\cite{rf:Zassenhaus}, in two dimensions. Here especially we are interested in superconducting films. 

$\!\!\!\!\!$ Superconducting films are made by repeated small increments of materials onto substrates held at low temperatures in an ultra-high vacuum. Then the system is found to form superconducting islands separated by thin insulating regions. The films are strongly disordered due to distribution of the island size and the coupling between islands. We assume, in this paper, that the randomness in island potential is crucial in determination whether the film is superconducting or insulating. 

$\!\!\!\!$ Natural questions arises : how much disorder is necessary to destroy the superconducting properties? Precise answer to this question will depend on details of the interaction among the particles, as well as the nature of the disorder in the system. 

Destruction of superconductivity with disorder can be caused by localization effect rather than Cooper pair braking. Then the system becomes insulator on the macroscopic scale in spite of local superconducting correlations. Relying upon a universality hypothesis, we regard Cooper pair as a tightly bound hard-core boson as long as we restrict ourselves to the vicinity of S-I transition point, disregarding any microscopic detail such as electrons, phonons, and their interactions. After all our problem becomes equivalent to dirty boson problem~\cite{rf:Fisher,rf:Herbut1,rf:Herbut2}, which has been extensively studied using quantum Monte Carlo simulations~\cite{rf:Scalettar,rf:Trivedi}, real-space renormalization-group techniques~\cite{rf:Zhang,rf:Singh}, strong-coupling expansions~\cite{rf:Freericks}, and other ways. 

$\!\!\!\!$ In this paper we shall adopt Matsubara-Matsuda model~\cite{rf:Matsubara,rf:Matsuda,rf:Liu} which was established for tightly bound hard-core bosons, and calculate the critical value of disorder which corresponds to the transition point with double time Green's function~\cite{rf:Zubarev,rf:Kondo,rf:Callen}. 

\section{ MODEL}
\label{SR}
\quad Inspired by the work of Matsubara and Matsuda (M-M) who formulated the model of hard-core boson in lattice gas~\cite{rf:Matsubara,rf:Matsuda,rf:Liu}, we introduce Bose-Hubbard type Hamiltonian :

\begin{equation}
{\cal H}=U\sum_{i,j}\hat{n}_i\hat{n}_j-t\sum_i(\hat{a}_i^{\dagger}\hat{a}_{i+1}+\hat{a}_i\hat{a}_{i+1}^{\dagger})-\sum_i(h+\delta h_i)\hat{n}_i, \label{eq:1}
\end{equation}
where $\hat{a}_i^{\dagger}$, $\hat{a}_i$, and $\hat{n}_i$ are the creation, annihilation, and number operators of Cooper pairs at the $i$-th lattice point, respectively. We exclude boson double occupancy by assuming infinite repulsive on-site interaction. We have introduced a finite repulsive interaction $U$, for two nearest-neighbor bosons. In addition, we restrict hopping $t$ to nearest-neighbor only. In this model we introduce random site potential $\delta h_i$ which represents disorder due to distribution of island size. It takes both positive and negative value. The external potential is described as the sum of uniform part $h$ and random part $\delta h_i$. The boson occupancy is restricted to one or zero. In other words each site has only two states, empty or full. The hardcore boson system is therefore isomorphic to a spin 1/2 system. Each operator of Cooper pairs can be replaced as spin operator~\cite{rf:Anderson} :
\begin{equation}
\hat{a}_i^{\dagger}\equiv\hat{S}_i^-\equiv\frac{1}{2}(\hat{\sigma}_i^x-i \hat{\sigma}_i^y),　\hat{a}_i\equiv\hat{S}_i^+\equiv\frac{1}{2}(\hat{\sigma}_i^x+i \hat{\sigma}_i^y) \label{eq:2},
\end{equation}

\begin{equation}
\hat{n}_i\equiv\frac{1}{2}-\hat{S}_i^z\equiv\frac{1}{2}(1-\hat{\sigma}_i^z) \label{eq:3},
\end{equation}
where $\hat{\sigma}_i^x$, $\hat{\sigma}_i^y$, and $\hat{\sigma}_i^z$ are Pauli matrices for the $1/2$ spin at the $i$-th lattice point, respectively. From (\ref{eq:1}), (\ref{eq:2}), and (\ref{eq:3}), the Hamiltonian becomes

\begin{equation}
{\cal H}=-J_z\sum_i \hat{S}_i^z\hat{S}_{i+1}^z-\frac{J_{xy}}{2}\sum_i(\hat{S}_i^+\hat{S}_{i+1}^-+\hat{S}_i^-\hat{S}_{i+1}^+)-\sum_i h_i\hat{S}_i^z \label{eq:4},
\end{equation}

\begin{equation}
J_z\equiv-4U ,　J_{xy}\equiv2t\label{eq:5},
\end{equation}

\begin{equation}
h_i\equiv h+\delta h_i\label{eq:6}.
\end{equation}
As a result, interaction energy and kinetic energy are now represented by spin exchange energy, and chemical potential is represented by Zeeman energy. Our system can be described with an anisotropic ferromagnetic Heisenberg Hamiltonian. 

$\!\!\!\!\!$ Therefore we can interpret ordering of pseudo-spins in $x$-$y$ plane as superconductivity and $z$ direction as local density fluctuation~\cite{rf:Matsubara}. 

\begin{equation}
\langle\hat{a}^{\dagger}\rangle , \langle\hat{a}\rangle\Leftrightarrow\langle\hat{S}^-\rangle , \langle\hat{S}^+\rangle \label{eq:7},
\end{equation}

\begin{equation}
\langle\hat{n}\rangle\Leftrightarrow\langle\hat{S}^z\rangle \label{eq:8}.
\end{equation}

$\!\!\!\!\!$ Since $\langle\hat{S}^+\rangle$ and $\langle\hat{S}^-\rangle$ represent superconducting order, their value can be regarded as a criterion to judge whether the system is, superconducting or insulating. So we shall calculate them in the next section.

\section{ CALCULATION}
\label{da}
\quad The problem now involves statistical physics of the pseudo-spin system described by the ferromagnetic Heisenberg model. The temperature dependent retarded Green's function~\cite{rf:Zubarev,rf:Kondo,rf:Callen} with two operators $\hat{S}_i^+$, $\hat{S}_j^-$ and a coefficient $h_i$, is introduced as

\begin{equation}
\langle\langle h_i\hat{S}_i^+(t);\hat{S}_j^-(0)\rangle\rangle\equiv-i\theta(t)\langle[h_i\hat{S}_i^+(t),\hat{S}_j^-(0)]\rangle. \label{eq:9}
\end{equation}
$\hat{S}_i^+(t)$ is a Heisenberg operator at time $t$ ; $\theta(t)$ is the step function ; square brackets $[\cdots]$ denote commutators, and single angular brackets $<\cdots>$ denote thermal averages. A straightforward calculation yields the time-Fourier transformed equation of motion for the Green's function (9)

\begin{equation}
\omega\langle\langle h_i\hat{S}_i^+;\hat{S}_j^-\rangle\rangle=\langle[h_i\hat{S}_i^+,\hat{S}_j^-]\rangle+\langle\langle[h_i\hat{S}_i^+,{\cal H}];\hat{S}_j^-\rangle\rangle. \label{eq:10}
\end{equation}

$\!\!\!\!$ In our two-dimensional square lattice model, we only consider coupling between nearest-neighbor sites (see Fig.1), i.e. site $i$ with site $i+a_n$($n=1\simeq 4$). 

\begin{eqnarray}
\omega\langle\langle h_i\hat{S}_i^+;\hat{S}_j^-\rangle\rangle&=&\langle[h_i\hat{S}_i^+,\hat{S}_j^-]\rangle+\langle\langle h_i^2\hat{S}_i^+;\hat{S}_j^-\rangle\rangle \nonumber \\ &&+\sum_a(J_z\langle\langle h_i\hat{S}_i^+\hat{S}_{i+a}^z;\hat{S}_j^-\rangle\rangle-J_{xy}\langle\langle h_i\hat{S}_i^z\hat{S}_{i+a}^+;\hat{S}_j^-\rangle\rangle). \label{eq:11}
\end{eqnarray}

$\!\!\!\!$ We can solve the equation of motion (\ref{eq:11}) easily by recovering translation invariance by impurity averaging procedure, and this can be done by averaging the magnetic field as follows 

\begin{eqnarray}
\langle h_i\rangle&=&\langle h+\delta h_i\rangle=h, \label{eq:12} \\
\langle h_i^2\rangle&=&\langle h^2+2h\delta h_i+\delta h_i^2\rangle=h^2+\delta h^2. \label{eq:13}
\end{eqnarray}
Since $\delta h_i$ takes both positive and negative values, the first order average equals to zero. 
Then the equation with translation invariance and randomness contribution is obtained as 

\begin{eqnarray}
\omega h\langle\langle\hat{S}_i^+;\hat{S}_j^-\rangle\rangle&=&h\langle[\hat{S}_i^+,\hat{S}_j^-]\rangle+(h^2+\delta h^2)\langle\langle\hat{S}_i^+;\hat{S}_j^-\rangle\rangle \nonumber \\
&&+\sum_a\left(J_z h\langle\langle\hat{S}_i^+\hat{S}_{i+a}^z;\hat{S}_j^-\rangle\rangle-J_{xy} h\langle\langle\hat{S}_i^z\hat{S}_{i+a}^+;\hat{S}_j^-\rangle\rangle\right). \label{eq:14}
\end{eqnarray}

$\!\!\!\!\!$ For simplicity, we decouple higher order Green's function in the following manner

\begin{equation}
\langle\langle\hat{S}_i^+\hat{S}_{i+a}^z;\hat{S}_j^z\rangle\rangle\rightarrow\langle\hat{S}^z\rangle\langle\langle\hat{S}_i^+;\hat{S}_j^z\rangle\rangle, \label{eq:15}
\end{equation}

\begin{equation}
\langle\langle\hat{S}_i^z\hat{S}_{i+a}^+;\hat{S}_j^z\rangle\rangle\rightarrow\langle\hat{S}^z\rangle\langle\langle\hat{S}_{i+a}^+;\hat{S}_j^z\rangle\rangle. \label{eq:16}
\end{equation}
Then (\ref{eq:14}) becomes

\begin{eqnarray}
\omega\langle\langle\hat{S}_i^+;\hat{S}_j^-\rangle\rangle&=&2\langle\hat{S}^z\rangle+\frac{h^2+\delta h^2}{h}\langle\langle\hat{S}_i^+;\hat{S}_j^-\rangle\rangle \nonumber \\
&+&\sum_a \left(J_z\langle\hat{S}^z\rangle\langle\langle\hat{S}_i^+;\hat{S}_j^-\rangle\rangle-J_{xy}\langle\hat{S}^z\rangle\langle\langle\hat{S}_{i+a};\hat{S}_j^-\rangle\rangle\right). \label{eq:17}
\end{eqnarray}
Fourier transformation (\ref{eq:17}) with respect to space results 

\begin{eqnarray}
\omega\langle\langle\hat{S}_k^+;\hat{S}_{-k}^-\rangle\rangle &=& 2\langle\hat{S}^z\rangle+\frac{h^2+\delta h^2}{h}\langle\langle\hat{S}_k^+;\hat{S}_{-k}^-\rangle\rangle \nonumber \\ &&+4\langle\hat{S}^z\rangle\left(J_z-\frac{J_{xy}}{4}\sum_a {\rm e}^{ika}\right)\langle\langle\hat{S}_k^+;\hat{S}_{-k}^-\rangle\rangle. \label{eq:18}
\end{eqnarray}
Putting them in order, we obtain the formula for the Green's function $\langle\langle\hat{S}_k^+;\hat{S}_{-k}^-\rangle\rangle$ 

\begin{equation}
G(k,\omega)\equiv\langle\langle\hat{S}_k^+;\hat{S}_{-k}^-\rangle\rangle=\frac{2\langle\hat{S}^z\rangle}{\omega-4\langle\hat{S}^z\rangle\omega_k-H}, \label{eq:19}
\end{equation}
where

\begin{equation}
\omega_k\equiv J_z-\frac{J_{xy}}{4}\sum_a {\rm e}^{ika}, \label{eq:20}
\end{equation}

\begin{equation}
H\equiv\frac{h^2+\delta h^2}{h}. \label{eq:21}
\end{equation}
Since the poles of Green's function indicates the excitations of the ferromagnetic spin wave, we see that two-body interaction $J_z$, external field $h$, and $\Delta h$ assist crystallization and hopping integral $J_{xy}$ causes quantum fluctuation. 

$\!\!\!\!\!$ Next we treat the correlation function $\langle\hat{S}_j^-\hat{S}_i^+\rangle$. It can be calculated by using the spectral theorem 

\begin{equation}
\langle\hat{S}^-\hat{S}^+\rangle=\frac{1}{4\pi^2}\int_{-\pi}^\pi dk_x\int_{-\pi}^\pi dk_y\frac{i}{2\pi}\int_{-\infty}^{\infty}d\omega\left[\{G_k^{\rm R}(\omega)-G_k^{\rm A}(\omega)\}\frac{1}{{\rm e}^{\frac{\omega}{T}}-1}\right],\label{eq:22}
\end{equation}
where $G_k^{\rm R}(\omega)$, $G_k^{\rm A}(\omega)$ are retarded and advanced Green's functions. From equation (\ref{eq:19}) we obtain

\begin{equation}
G_k^R(\omega)-G_k^A(\omega)=-2\pi i\delta(\omega-4\langle\hat{S}^z\rangle\omega_k-H),\label{eq:23}
\end{equation}
then the integration of the right-hand side of (\ref{eq:22}) with respect to $\omega$ becomes

\begin{equation}
\frac{1}{2}-\langle\hat{S}^z\rangle=\frac{1}{4\pi^2}\int_{-\pi}^\pi {\rm d}k_x\int_{-\pi}^\pi {\rm d}k_y\frac{2\langle\hat{S}^z\rangle}{{\rm exp}[(4\langle\hat{S}^z\rangle\omega_k+H)/(2T)]-1}. \label{eq:24}
\end{equation}
Putting them in order, we obtain the relation 

\begin{equation}
1=\frac{\langle\hat{S}^z\rangle}{\pi}\int_{0}^\pi k{\rm d}k {\rm coth}\left(\frac{4\langle\hat{S}^z\rangle\omega_k+H}{2T}\right). \label{eq:25}
\end{equation}
Since low temperature region is of our main interest, we neglect large spin wave vector. This approximation gives $\omega_k$ as 

\begin{equation}
\omega_k\simeq\frac{1}{4}J_{xy}k^2+J_z-J_{xy}.\label{eq:26}
\end{equation}
Therefore the excitation energy gives 

\begin{equation}
\omega=\langle\hat{S}^z\rangle J_{xy}k^2+4\langle\hat{S}^z\rangle(J_z-J_{xy})+H.\label{eq:27}
\end{equation}
The second and third term on the right-hand side means the gap of the spin wave excitation $\Delta$ at $k\simeq 0$, which is caused by anisotropy and external field. Here two conditions must be fullfilled in order that the ground state is superconducting. The first one is that $\Delta$ is sufficiently small compared to $T$, i.e. 

\begin{equation}
\Delta=4\langle\hat{S}^z\rangle(J_z-J_{xy})+H\ll T, \label{eq:28}
\end{equation}
the second one is 

\begin{equation}
J_{xy}> J_z. \label{eq:29}
\end{equation}
Using the above conditions, the integration on the right-hand side of (\ref{eq:25}) can be calculated analytically, and so we obtain the expression at low temperature 

\begin{equation}
\langle\hat{S}^z\rangle=\frac{H(\rm{e}^{\pi J_{xy}/T}-1)}{\pi^2 J_{xy}-4(J_z-J_{xy})(\rm{e}^{\pi J_{xy}/T}-1)}\simeq\frac{H}{4(J_z-J_{xy})}\label{eq:30}.
\end{equation}
Due to rotation symetry of spin in $x$-$y$ plane, we can select the $x$-axis as the grand state direction without lost of generality. Then we obtain the relation between the superconducting order parameter $\langle\hat{S}^x\rangle$ and randomness $\Delta h$ 

\begin{equation}
\langle\hat{S}^x\rangle=\sqrt{\left(\frac{1}{2}\right)^2-\langle\hat{S}^z\rangle^2}\simeq\frac{1}{2}\sqrt{1-\frac{H^2}{4(J_z-J_{xy})^2}}.\label{eq:31}
\end{equation}

\section{ RESULT}
\label{aa}
\quad Before examining the physical meanings of (\ref{eq:31}), we introduce some normalized factor

\begin{equation}
\eta\equiv\frac{\Delta h}{J_{xy}},　\kappa\equiv\frac{J_z}{J_{xy}},　\xi\equiv\frac{h}{J_{xy}}.\label{eq:32}
\end{equation}
Which corresponds to normalized strength of randomness, two body interaction, and chemical potential, respectively. Then (\ref{eq:31}) becomes 

\begin{equation}
\langle\hat{S}^x\rangle\simeq\frac{1}{2}\sqrt{1-\frac{(\xi^2+\eta^2)^2}{4\xi^2(1-\kappa)^2}}.\label{eq:33}
\end{equation}
This equation describes the behavior of the superconducting order parameter $\langle\hat{S}^x\rangle$ with respect to the strength of randomness $\eta$ (see Fig.2).
From Fig.2 we can see that $\langle\hat{S}^x\rangle$ is damped by increasing $\eta$. By expanding (\ref{eq:33}) in power of $\eta$ 

\begin{equation}
\langle\hat{S}^x\rangle\simeq\frac{1}{2}\sqrt{1-\frac{\xi^2}{4(1-\kappa)^2}}\left[1-\frac{\eta^2}{4(1-\kappa)^2-\xi^2}\right],\label{eq:34}
\end{equation}
we can see the damping of $\langle\hat{S}^x\rangle$ occurs proportional to $\eta^2$. At a specific point, $\langle\hat{S}^x\rangle$ completely disappears, which implies the existence of a phase transition. As mentioned in $\sec.2$, we interpret phases $\langle\hat{S}^x\rangle\neq 0$ as superconducting and $\langle\hat{S}^x\rangle=0$ as insulating. Here the central result is the existence of the critical value $\eta_c$ for the strength of randomness. $\eta_c$ divides the two phases (S and I) and gives the transition point. From (\ref{eq:33}) $\eta_c$ is given by

\begin{equation}
\eta_c=\sqrt{2\xi(1-\kappa)-\xi^2}.\label{eq:35}
\end{equation}
From Fig.2 we can see that the strength of two-body interaction $\kappa$ advances the transition. Therefore $\kappa$ can be also regarded as a parameter of S-I transition. (see also Fig.3). 

$\!\!\!\!\!$ Varying $\kappa$ instead of $\eta$ (see Fig.3) also gives a phase transition, this time it corresponds to a Mott transition. Especially when $\eta=0$, $\langle\hat{S}^x\rangle$ sharply decreases at $\kappa\simeq 1$. This corresponds to a transition in an 2D isotropic system i.e. $\eta=0$, $\kappa=1$. This is based on the general fact that $\langle\hat{S}^x\rangle$ doesn't exist in two-dimensional system. 

$\!\!\!\!\!$ Now we vary both $\eta$ and $\kappa$ as a parameter of the phase transition, the border line between two phases (S and I) is determined by the condition that superconductivity disappears (i.e. $\langle\hat{S}^x\rangle=0$). 

$\!\!\!\!$ From (\ref{eq:33}) this relation corresponds to
\begin{equation}
1-\frac{(\xi^2+\eta^2)^2}{4\xi^2(1-\kappa)^2}=0. \label{eq:36}
\end{equation}
Real $\langle\hat{S}^x\rangle$ corresponds to the system being superconducting, in this case $\eta$, $\kappa$ have to satisfy 

\begin{equation}
1-\frac{(\xi^2+\eta^2)^2}{4\xi^2(1-\kappa)^2}>0. \label{eq:37}
\end{equation}
On the other hand, non-real $\langle\hat{S}^x\rangle$ corresponds to the system being insulating, this time $\eta$, $\kappa$ have to satisfy

\begin{equation}
1-\frac{(\xi^2+\eta^2)^2}{4\xi^2(1-\kappa)^2}<0. \label{eq:38}
\end{equation}
Then we can draw a phase diagram of S-I transition with respect to $\eta$ and $\kappa$ (see Fig.4). 

$\!\!\!\!\!\!$ (\ref{eq:36}) $\sim$ (\ref{eq:38}) can be also expressed as follows 

\begin{equation}
J_{xy}=J_z+\frac{1}{2}\left(h+\frac{\Delta h^2}{h}\right) 　　(\rm{critical}\;\;\rm{region}),　　　　\label{eq:39}
\end{equation} 

\begin{equation}
J_{xy}>J_z+\frac{1}{2}\left(h+\frac{\Delta h^2}{h}\right) 　　(\rm{superconducting}\;\;\rm{region}), \label{eq:40}
\end{equation}

\begin{equation}
J_{xy}<J_z+\frac{1}{2}\left(h+\frac{\Delta h^2}{h}\right) 　　(\rm{insulating}\;\;\rm{region}). \;\;\;\,　　\label{eq:41}
\end{equation}
The left-hand side is kinetic energy which causes quantum fluctuation. The right-hand side is potential energy which causes crystallization. In other words the former indicates wave property and the latter indicates particle property (localization) of Cooper pair, Therefore the phase diagram of our system is determined by the competition between them. 

\section{ SUMMARY}
\label{td}
\quad We have focused on the S-I transition in ultra-thin films and considered randomness contribution as a factor to determine the border between the two phases (S and I). Superconducting order parameter $\langle\hat{S}^x\rangle$ has been calculated as function of randomness $\eta$, and influence of $\eta$ to S-I transition has been investigated. As a result, we have shown that $\langle\hat{S}^x\rangle$ decreases with increasing $\eta$ and obtained analytically the critical value of randomness $\eta_c$. 

$\!\!\!$ The existence of random potential causes scattering of Cooper pairs, and disturbs their free motion. Randomness $\eta$ becomes an important factor for the occurrence of electric resistance $R$. There is a one-to-one correspondence between the critical value of $R$ and $\eta$ ($R_c$ and $\eta_c$). Therefore solving $\eta_c$ is equivalent to solving $R_c$. It has been well known that the value of critical sheet resistance in ultrathin films is on the order of $h/4e^2$, but its exact coefficients has not been calculated yet. Our results may have some relevance to the determination of $R_c$. 

$\!\!\!\!\!\!$ We have obtained the results that randomness and two-body interaction causes localization and destroys superconductivity. These are capable of explaining, at least qualitatively, the main characteristic of S-I transition. In order to proceed to a more quantitative discussion, it is necessary to employ a more refined approximation in evaluating higher-order Green's functions. 

\acknowledgments
\quad We are very grateful to S. Saito, K. Sano, and B. H. Valtan for useful discussions and critical reading.

\begin{figure}
\caption{ Two-dimensional square lattice. The four sites $i+a_n$ $(n=1\sim4)$ are the nearest-neighbor sites of the site $i$ on the square lattice. }\label{f1}
\end{figure}

\begin{figure}
\caption{ The behavior of superconductivity $\langle\hat{S}^x\rangle$ with increasing randomness $\eta$. $\langle\hat{S}^x\rangle$ continuously decreases and completely disappears at each specific point. Two body interaction $\kappa$ contributes to advancing the transition from superconducting to insulating. } \label{f5}
\end{figure}
\begin{figure}
\caption{ The behavior of superconductivity $\langle\hat{S}^x\rangle$ with increasing two body interaction $\kappa$. $\langle\hat{S}^x\rangle$ continuously decreased and completely disappeared at each specific point. Randomness $\eta$ contributes to advancing the transition from superconducting to insulating. } \label{f10}
\end{figure}

\begin{figure}
\caption{The phase diagram for our lattice boson model with randomness $\eta$ and two-body interaction $\kappa$. The borderline between two phases (S and I) is determined by the competition between wave property and particle property. When wave property is stronger than particle property, the system is in superconducting. When particle property is stronger than wave property, the system is in insulating. } \label{faaz}
\end{figure}

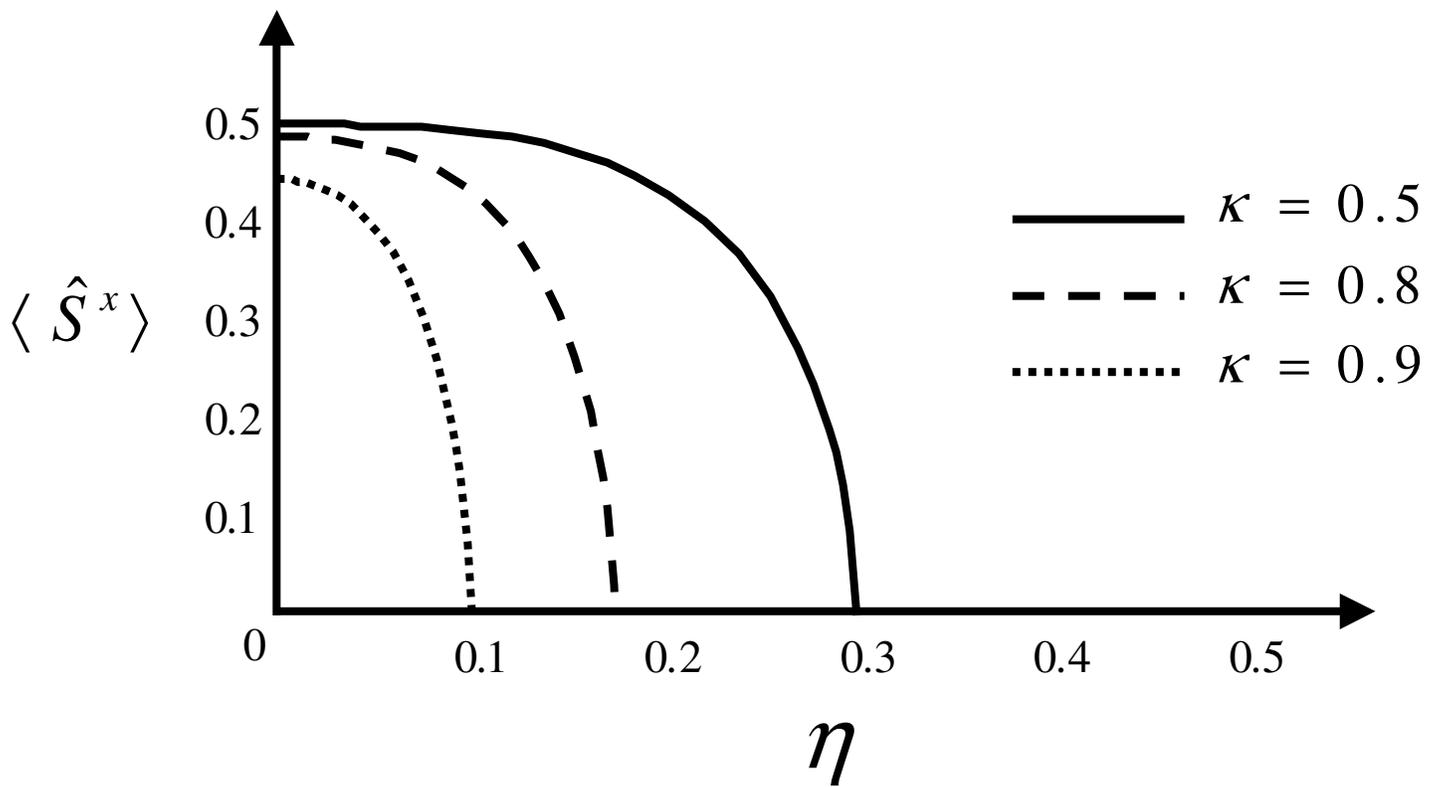
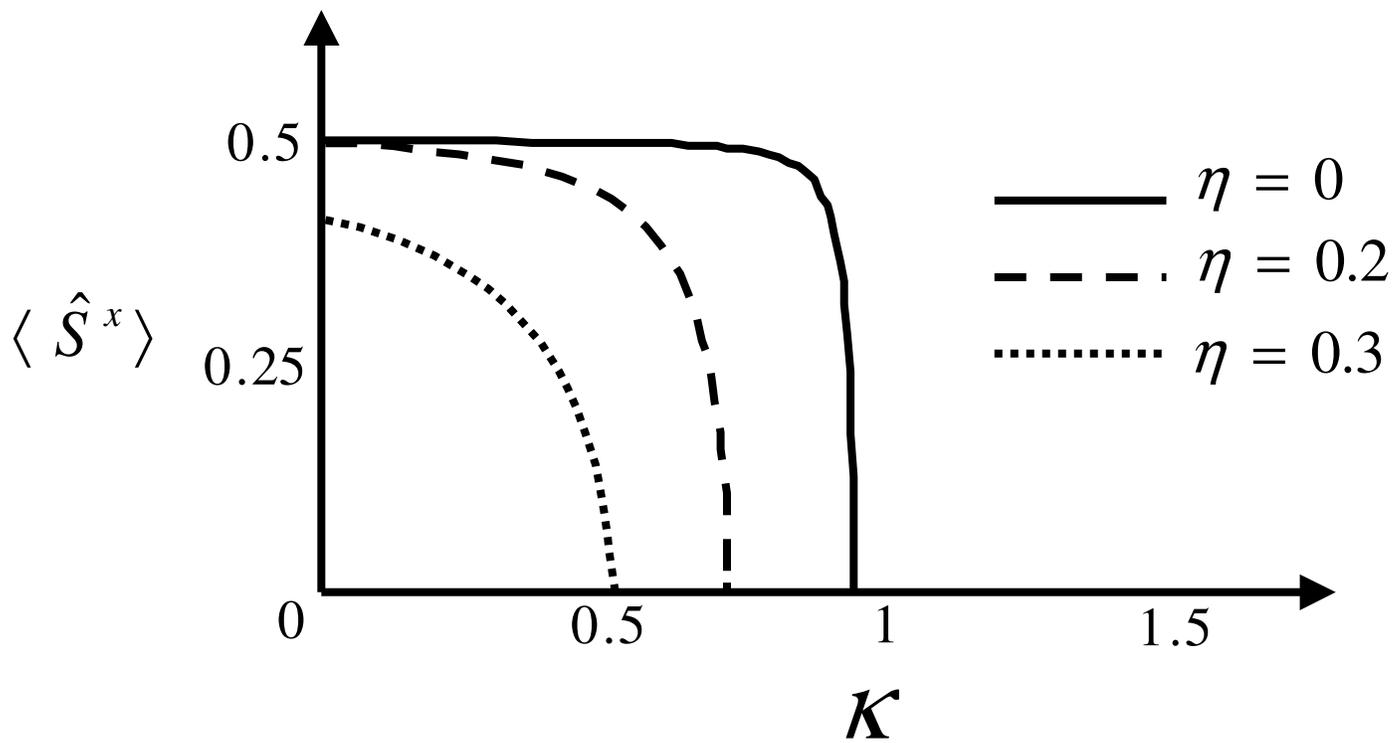
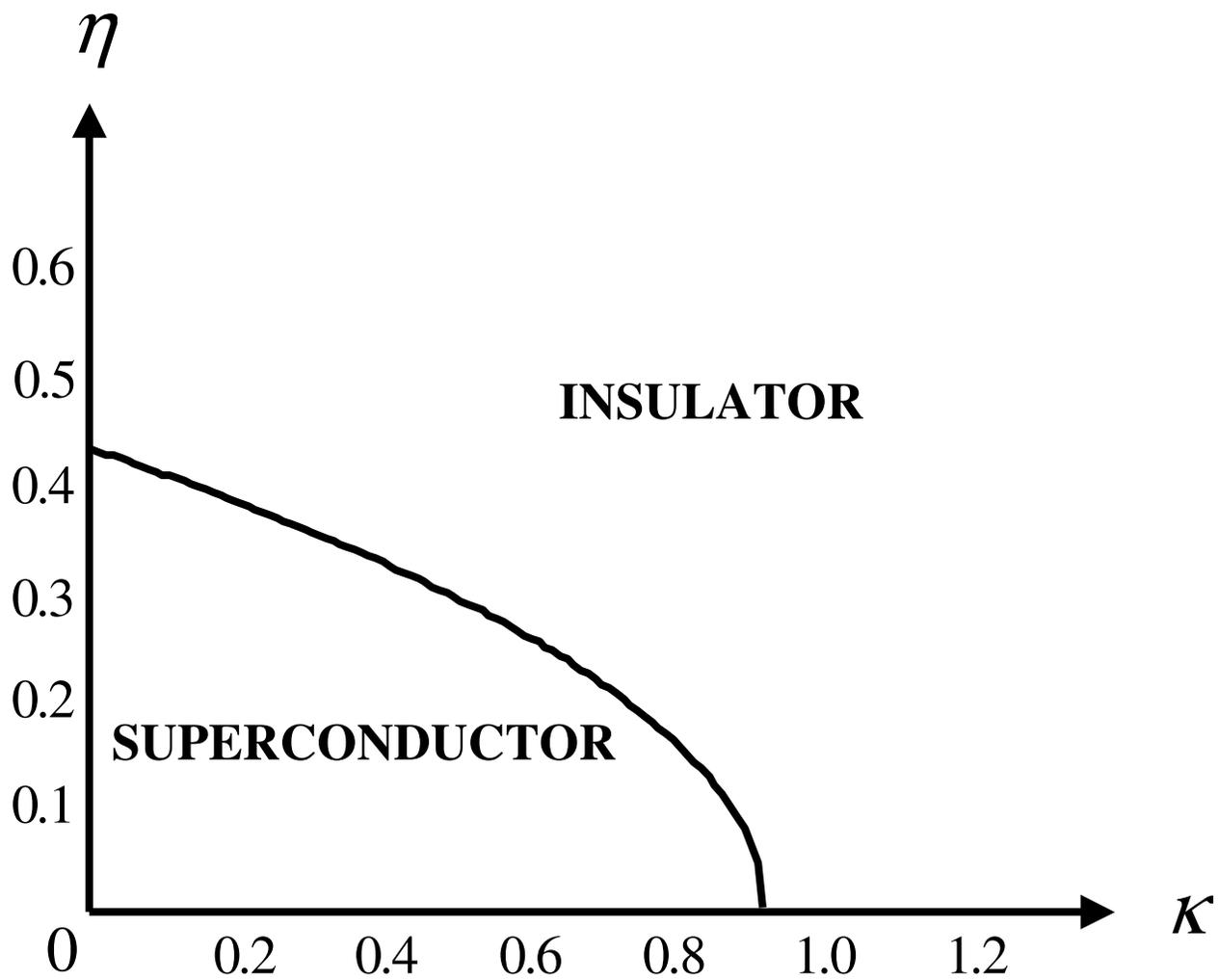
\end{document}